\documentclass{ifacconf}

\usepackage{graphicx}      %
\usepackage{natbib}        %
\usepackage{fancyhdr}
\makeatletter
\newcommand*\titleheader[1]{\gdef\@titleheader{#1}}
\AtBeginDocument{%
  \let\st@red@title\@title
  \def\@title{%
    \bgroup\normalfont\large\centering\@titleheader\par\egroup
    \vskip1.5em\st@red@title}
}
\makeatother

\usepackage{amssymb,amsmath,bm}
\usepackage{graphicx,float}
\newcommand{\beq}{\begin{equation}}
\newcommand{\eeq}{\end{equation}}

\usepackage{algorithm}
\usepackage{algpseudocode}
\begin{document}

\begin{frontmatter}
\vspace{-2em}
© 2024 Artur Wolek. This work has been accepted to IFAC for publication under a Creative Commons Licence CC-BY-NC-ND
\title{Path Planning for a Cooperative Navigation Aid Vehicle to Assist Multiple Agents Sequentially}

\author[First]{Artur Wolek} 

\address[First]{Department of Mechanical Engineering and Engineering Science, \\ University of North Carolina at Charlotte, Charlotte, NC, 28223 USA (e-mail: awolek@charlotte.edu).}

\begin{abstract}
This paper considers planning a path for a single underwater cooperative navigation aid (CNA) vehicle to sequentially aid a set of $N$ agents to minimize average navigation uncertainty. Both the CNA and agents are modeled as constant-velocity vehicles. The agents travel along known nominal trajectories and the CNA plans a path to sequentially intercept them. Navigation aiding is modeled by a scalar discrete time Kalman filter. During path planning, the CNA considers surfacing to reduce its own navigation uncertainty. A greedy planning algorithm is proposed that uses a heuristic to schedule agents to the CNA that is based on the optimal time-to-aid, the overall navigation uncertainty reduction, and the transit time. The approach is compared to an optimal (exhaustive enumeration) algorithm through a Monte Carlo experiment with randomized agent  trajectories and initial navigation uncertainty.
\end{abstract}
\begin{keyword}
cooperative navigation, underwater vehicles, path planning
\end{keyword}

\end{frontmatter}

\section{Introduction}

A cooperative navigation aid (CNA) is a surface or underwater vehicle that is equipped with a high-performance navigation system that serves as a moving reference position transponder and provides relative range and/or bearing measurements to other nearby vehicles with less capable navigation systems \citep{Vagnay.OES.2004}. Prior work on path planning for CNAs has focused on close-proximity and continuous aiding scenarios (i.e., assuming the CNA remains within communication range of the vehicles being aided at all times). The CNA-to-agent geometry during these interactions determines the state estimate and navigation uncertainty for each agent receiving the aiding. Optimal steering and path planning strategies to improve localization performance during such interactions have been explored. For example,  \cite{Quenzer2014} used a helming strategy guided by path inertia or the empirical observability Gramian. \cite{chitre2010path} used approximate dynamic programming principles with a value function related to the sum-square position error of agents. The problem has also been formulated as a Markov decision process with cross-entropy learning  \citep{Teck.OCEANS.2011}.  \cite{Bahr2009} demonstrated a CNA path optimization strategy experimentally. 

In this work, a related scenario is considered wherein the CNA  can provide \emph{sequential} aiding to multiple agents. That is, we assume that there are $N$ agents requiring navigation assistance, but the CNA can only service one at a time due to their spatial separation. The  goal is to design a path that minimizes average navigation uncertainty across all agents by optimizing the sequence in which all agents (or a subset of agents) are intercepted subject to a time constraint. The optimization also considers the possibility that the CNA can reduce  its own navigation uncertainty by surfacing for a GPS fix. We envision that the proposed method can be used as a high-level task scheduling algorithm to dectermine the sequence of agents to aid, and existing steering strategies can optimize close-range interactions to improve localization performance. 

%This paper considers the path planning problem for a single CNA to sequentially aid a set of $N$ agents to minimize average navigation uncertainty with a mission time constraint. 
The contributions of this paper are: (1) a derivation of the optimal time-to-aid for a CNA aiding a single agent, and (2) a greedy algorithm that uses a novel heuristic to efficient solve the sequential aiding problem. The performance of the proposed algorithm is evaluated through Monte Carlo simulations.
% and compared to an optimal (exhaustive enumeration) algorithm.

The remainder of the paper is organized as follows. Section~\ref{sec:problemStatement} describes the environment, CNA and agent motion, the navigation uncertainty models, and the cost function to be optimized. 
Section~\ref{eq:single_contact} presents intermediate results for a single CNA and agent scenario. Section~\ref{sec:optimization_methods} introduces the greedy algorithm. Section~\ref{sec:results} presents the results of numerical simulations, and the paper is concluded in Sec.~\ref{sec:conclusion}.

\section{Problem Formulation}
\label{sec:problemStatement}
This work considers the problem of planning the path of a cooperative navigation aid (CNA) to aid $N$ other vehicles (referred to as agents) traveling along known nominal trajectories with known initial navigational uncertainty.  In this section the environment model, vehicle motion models, navigation uncertainty models are described and the problem statement is introduced.
\subsection{Environment and CNA and Agent Motion Models}
The CNA and agents are assumed to operate in the plane, and 
the $i$th agent's state at discrete time $k$ is the position $\bm{x}^i_k = [x^i_k,y^i_k]^{\rm T}$. The agent's motion is described by the  constant-velocity dynamics 
\begin{equation}
\begin{split}
{\bm x}_k^i &= {\bm F}{\bm x}_{k-1}^i + {\bm G}{\bm u}_{k-1}^i + {\bm w}_k \\
&=
    \left[
    \begin{array}{cc}
    1 & 0 \\
    0 & 1 \\
    \end{array}
    \right]
    \left[
    \begin{array}{c}
    x^i_{k-1} \\
    y^i_{k-1} \\
    \end{array}
    \right]+
    \left[
    \begin{array}{cc}
    1 & 0 \\
    0 & 1 \\
    \end{array}
    \right]
    \left[
    \begin{array}{c}
    u^i_{x,k-1} \\
    u^i_{y,k-1} \\
    \end{array}
    \right]+
    {\bm w}_k\;,    
\end{split}
\label{eq: Agent Dynamics}
\end{equation}
where  ${\bm u}_{k-1}^i = [u_{x,k-1}^i,  u_{y,k-1}^i]^{\rm T}$ is a velocity control input constrained to have magnitude $||{\bm u}_{k-1}^i||  = v_a \Delta t$, $v_a$ is the agent's speed, and $\Delta t = t_k - t_{k-1}$ is the timestep of the discretized motion model. The agent dynamics include zero-mean Gaussian process noise ${\bm w}_k \sim \mathcal{N}({\bm 0}, {\bm Q})$ where ${\bm Q} = {\bm 1}\nu_w$ is a diagonal matrix with entries of variance $\nu_w$ and ${\bm 1}$ is the identity matrix.

Similarly, the CNA's state is the position $\bm{x}^c_k = [x^c_k,y^c_k]^{\rm T}$, and the CNA's motion is described by the dynamics
\begin{equation}
\begin{split}
{\bm x}_k^c &= {\bm F}^c{\bm x}_{k-1}^c + {\bm G}^c{\bm u}_{k-1}^c + \bm{w}^c_k \\
&=
    \left[
    \begin{array}{cc}
    1 & 0 \\
    0 & 1 \\
    \end{array}
    \right]
    \left[
    \begin{array}{c}
    x^c_{k-1} \\
    y^c_{k-1} \\
    \end{array}
    \right]+
    \left[
    \begin{array}{cc}
    1 & 0 \\
    0 & 1 \\
    \end{array}
    \right]
    \left[
    \begin{array}{c}
    u^c_{x,k-1} \\
    u^c_{y,k-1} \\
    \end{array}
    \right]+
    \bm{w}^c_k\;,
    \label{eq: CNA Dynamics}
\end{split}
\end{equation}
where  ${\bm u}_{k-1}^c = [u_{x,k-1}^c,  u_{y,k-1}^c]^{\rm T}$ is a velocity control input constrained to have magnitude $||{\bm u}_{k-1}^c||  \leq v_c \Delta t$,  and $ v_c$ is the CNA's speed. Let $\eta = v_a/v_c < 1$ denote a speed ratio. The CNA's dynamics include zero-mean Gaussian process noise ${\bm w}^c_k \sim \mathcal{N}({\bm 0}, {\bm Q}^c)$ where ${\bm Q}^c = {\bm 1}\nu_c$ is a diagonal matrix with entries of  variance $\nu_c$.  The process noise terms model the effects of navigation uncertainty (i.e., the position variables ${\bm x}^c$ and ${\bm x}^i$ represent the time-evolution of the navigation system's estimate). The CNA is assumed to have a more capable navigation system, thus $\nu_c \ll \nu_w$.

\subsection{Agent Navigation Uncertainty Model}
\label{sec:nav_error_agent}
Each agent begins at an uncertain initial position ${\bm x}_0^i \sim \mathcal{N}(\hat {\bm x}_{0|0}^i, {\bm P}^i_{0|0})$ with mean $\hat {\bm x}_{0|0}^i = [x^i_{0|0},y^i_{0|0}]^{\rm T}$ and covariance ${\bm P}^i_{0|0} = {\bm 1}\nu^i_{0|0}$.
 The CNA provides absolute position  measurements to the $i$th agent at time $k$: 
\begin{equation}
\begin{split}
 {\bm y}_k^i &= {\bm H}{\bm x}_k^i + {\bm v}_k = \left[ 
\begin{array}{c}
x_k^i \\
y_k^i
\end{array}
\right]+ {\bm v}_k \;,
\label{eq:agent_measurement}
\end{split}
\end{equation}
where ${\bm H}$ is an identity matrix, ${\bm v}_k \sim \mathcal{N} ({\bm 0},{\bm R}_k)$ is zero-mean Gaussian measurement noise, and ${\bm R}_k = {\bm 1} \nu_y + {\bm P}_{k|k}^c$. 
The covariance of the measurement noise, ${\bm R}_k$, consists of a constant $\nu_y$ and a time-varying term that depends on the absolute position uncertainty of the CNA itself, ${\bm P}_{k|k}^c$.  
This assumption models the fact that as the CNA's own uncertainty increases so does the uncertainty in the aiding measurements it provides. Aiding is assumed to occur only when the agent and CNA are collocated.

If a measurement is available, the agent's state estimate is updated from the previous time  ($k-1$) according to the discrete-time Kalman filter equations \citep{simon2006optimal}:
\begin{align}
\hat{\bm{x}}^i_{k|k-1} &= \bm{F}\hat{\bm{x}}^i_{k-1|k-1}+\bm{G}\bm{u}^i_{k-1}\label{eq:State Prediction Agent} \\
\bm{P}^i_{k|k-1} &= \bm{F}\bm{P}^i_{k-1|k-1}\bm{F}^{\rm T} + \bm{Q} \label{eq:Covariance Prediction}\\
\bm{y}^i_{k|k-1} &= \bm{H}\hat{\bm{x}}^i_{k|k-1} \label{eq:Measurement Prediction}\\
\bm{S}_k &= \bm{R}_k + \bm{H}\bm{P}^i_{k|k-1}\bm{H}^{\rm T} \label{eq:Innovation Covariance}\\
\bm{K}_k &= \bm{P}^i_{k|k-1}\bm{H}^{\rm T}\bm{S}_k^{-1} \label{eq:Kalman Gain Agent}\\
\hat{\bm{x}}^i_{k|k} &= \hat{\bm{x}}^i_{k|k-1} + \bm{K}_k(\bm{y}^i_k - \bm{y}^i_{k|k-1})\label{eq:State Prosterior Agent}\\
\bm{P}^i_{k|k} &= (\bm{I} - \bm{K}_k\bm{H})\bm{P}^i_{k|k-1}\label{eq:Covariance Posterior Agent}\;,
\end{align}
where $\hat{\bm{x}}^i_{k-1|k-1}$ is the state prior, $\hat{\bm{x}}^i_{k|k-1}$ is the state prediction, $\hat{\bm{x}}^i_{k|k}$ is the state posterior, and the matrices ${\bm F}, {\bm G}, {\bm Q}, {\bm H}, {\bm R}$ are given in \eqref{eq: Agent Dynamics} and \eqref{eq:agent_measurement}. 
If, instead, the CNA is not collocated with the $i$th agent then no measurements are transmitted.
In these cases, the agent's own navigation estimate is updated as:
\begin{align}
\hat{\bm{x}}^i_{k|k} &= \bm{F}\hat{\bm{x}}^i_{k-1|k-1}+\bm{G}\bm{u}^i_{k-1}\label{eq:State Prediction Agent_nomsmt} \\
\bm{P}^i_{k|k} &= \bm{F}\bm{P}^i_{k-1|k-1}\bm{F}^{\rm T} + \bm{Q} \label{eq:Covariance Prediction_nomsmt} \;.
\end{align}
\subsection{CNA Navigation Uncertainty Model}
\label{sec:nav_error_cna}
The CNA begins at an uncertain initial position ${\bm x}_0^c \sim \mathcal{N}(\hat {\bm x}_{0|0}^c, {\bm P}^c_{0|0})$ with mean $\hat {\bm x}^c_{0|0}$ and covariance ${\bm P}^c_{0|0}={\bm 1}\nu_G$, where $\nu_G$ is a variance. Similar to \eqref{eq:State Prediction Agent_nomsmt}--\eqref{eq:Covariance Prediction_nomsmt}, the CNA's navigation uncertainty evolves according to:
\begin{align}
\hat{\bm{x}}^c_{k|k} &= \bm{F}\hat{\bm{x}}^c_{k-1|k-1}+\bm{G}\bm{u}^c_{k-1}\label{eq:CNA_mean_evolve} \\
\bm{P}^c_{k|k} &= \bm{F}\bm{P}^c_{k-1|k-1}\bm{F}^{\rm T} + \bm{Q}^c \label{eq:CNA_cov_evolve} \;.
\end{align}
However, the  CNA is capable of periodically surfacing to obtain a GPS fix. A CNA can surface at discrete time $S$ that lasts for $M$ timesteps. For simplicity, we assume the CNA is unavailable to provide navigation updates and does not move during this time (i.e., $\bm{u}^c_{j} = {\bm 0}$ for $j \in \{ S, S+1, ...S+M\}$). After $M$ timesteps have elapsed the navigation uncertainty is re-initialized as
\begin{align}
\hat{\bm{x}}^c_{S+M|S+M} &\sim \mathcal{N}({\bm{x}}^c_{k|k} , \bm{V} ) \\  
\bm{P}^c_{S+M|S+M} &=   {\bm 1} \nu_G \label{eq:CNA_cov_evolve} \;.
\end{align}
The variance $\nu_G$ is the same as the initial CNA variance and represents the uncertainty of the CNA after it has completed surfacing  and returned to an operating depth.  
\subsection{Problem Statement}\label{Section: Cost Function}
The goal is to design a path for the CNA to minimize average navigation uncertainty across all agents with a mission time constraint. Suppose that the CNA has prior knowledge that the agents are travelling from their nominal initial states, $\bm{\bar x}^i_0 = \hat {\bm x}^i_0$, to a final state, $\bm{ \bar x}_T^i$, along a known straight-line trajectory $\bm{ \bar x}_{0:T}^i = \{ \bm{ \bar x}_0^i, \bm {\bar x}_1^i, \ldots,  \bm{\bar x}_T^i \}$ where the final time   is $T$. The CNA begins at the initial state $\bar {\bm{x}}^c_0 = \hat {\bm x}^c_0$. The optimization problem is to determine the CNA trajectory $\bar {\bm x}_{0:T}^c = \{ \bar {\bm x}_0^c, {\bm x}_1^c, \ldots,  \bar {\bm x}_T^c \}$ and time to surface $S$ that 
\begin{equation}
{\rm minimize}~~J = \frac{1}{2N(T+1)} \sum^N_{i=1} \sum^{T}_{k=0}{\rm Trace}(\bm{P}^i_{k|k})\;,
\label{eq:prob}
\end{equation}
subject to the dynamics \eqref{eq: Agent Dynamics}--\eqref{eq: CNA Dynamics}, measurement model \eqref{eq:agent_measurement}, and navigation uncertainty models \eqref{eq:State Prediction Agent}--\eqref{eq:CNA_cov_evolve}. The cost $J$ in \eqref{eq:prob} is the average  navigation uncertainty across all agents.

To reduce complexity, the optimization is restricted to a class of CNA paths that are parameterized by an ordered sequence of at most $D \leq N+1$ unique  tasks. Task are from the set $\mathcal{Q} = \{ 0, 1, 2, \ldots, N\}$, where an integer task $a  > 0$ indicates intercepting the $i$th agent and the task $a = 0$ indicates surfacing. A $D$-length task sequence is denoted by ${A} = \{ a_1,\ldots, a_D \}$. The set of feasible task sequences is of length $D$ or less is:
\begin{align*}
\mathcal{F} = \{ 
&{A} =  \{a_1,\ldots, a_F \}~|~a_i \in \mathcal{Q} ~\text{for all}~ i=1, \ldots, F,  \notag \\
& ~F \leq D, ~ a_i \neq a_j ~\text{for all}~ i \neq j , ~ T({A}) \leq T_{\rm max} 
\}\;,
\end{align*}
where $T({A})$ is the time to complete ${A}$.  The following section describes the  strategy used by the CNA to intercept each agent. Using this strategy, a sequence ${A} \in \mathcal{F}$ is mapped to a  CNA path $\{ \bar {\bm x}_0^c, {\bm x}_1^c, \ldots,  \bar {\bm x}_T^c \}$ and cost  \eqref{eq:prob}. 
\section{Single-Agent Aiding Intercepts}
\label{eq:single_contact}
This section presents the strategy used to intercept an agent and the change in the cost function associated with such an interaction. These results are then used in Sec.~\ref{sec:optimization_methods} to design a combined task and path planning algorithm.
\subsection{Minimum Time to Intercept an Agent}
\begin{figure}[h!]
    \centering
    \includegraphics[width = 0.25\textwidth]{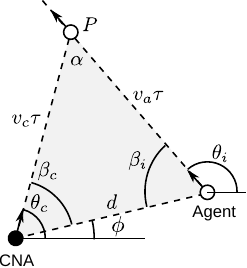}
    \caption{Geometry for computing the time to intercept agent $i$ by the CNA.}
    \label{fig:intercept}
\end{figure}

Suppose that at the initial time $t_0$ an agent is located at $(x_{i}(t_0), y_{i}(t_0))$ and traveling along a heading $\theta_i$ at a speed $v_a$. 
At time $t_0$ the CNA is located at $(x_c(t_0), y_c(t_0))$ and it seeks to select the heading angle $\theta_c$ to intercept the agent.  Consider a triangle formed by the agent's position, the CNA's position, and a future intersection point $P$. Let $\alpha$, $\beta_i$, and $\beta_c$ be the interior triangle angles formed by this triangle (as shown in Fig.~\ref{fig:intercept}), and $\tau$ is the time to intercept.

Define the angle $\phi_i = ~{\rm atan2}(y_i-y_c, x_i-x_c)$  from the CNA to the agent.  Let ${\bm \phi}_i = [\cos \phi_i, \sin \phi_i, 0]^{\rm T}$ be the vector pointing from the  CNA to the $i$th agent. Moreover, let ${\bm \theta}_i = [\cos \theta_i, \sin \theta_i,0]^{\rm T}$ be the vector pointing along the path of the $i$th agent. 
The internal angles are then: 
\begin{align*}
\beta_i = {\rm acos} ( \langle-{\bm \phi}_i ,  {\bm \theta_i} \rangle)  \quad \text{and} \quad 
\beta_c &= {\rm acos} ( \langle {\bm \phi}_i, {\bm \theta_c} \rangle)  \;,
\end{align*} 
where ${\bm \theta}_c = [\cos \theta_c, \sin \theta_c, 0]^{\rm T}$ is the travel direction for the CNA (to be determined).
Note that $\alpha = \pi - \beta_c - \beta_i$. From the sine law
 the intercept angle is 
\begin{equation*}
\beta_c = {\rm asin}\left( \eta  {\sin \beta_i}  \right) \;.
\end{equation*}
Moreover, if $d$ is the initial distance between the CNA and agent then the time to intercept is 
\begin{align}
\tau &= \frac{d \sin \beta_c}{v_a\sin\alpha}  %
\label{eq:intercept_time} \;.
\end{align}
If $\theta_i = \phi + \pi$, then the agent is moving towards the initial CNA position and $\theta_c = \phi$ 
 and $\tau = d/(v_a  + v_c)$. If $\theta_i = \phi $, then the agent is moving away from the CNA and 
$\tau = d/(v_c - v_a)$ and $\theta_c = \theta_i$. 
Otherwise, 
$
\theta_c =  \phi + \delta \beta_c
$\;, 
where $\delta = {\rm sign}(\delta_3)$ and $ {\bm \phi}_i \times {\bm \theta}_i = [\delta_1, \delta_2, \delta_3]^{\rm T}$.
\subsection{Agent Navigation Uncertainty after CNA Interaction}
Now, consider how the uncertainty of each agent evolves given that they are intercepted at some timestep $ 1 \leq Z \leq T$ and receive a single measurement update from the CNA.  Since the initial covariance of each agent and the agent state and input matrices  in \eqref{eq: Agent Dynamics} are diagonal, and equal in both directions, it follows that the covariance of the agent is characterized by a scalar variance $\nu^i_{k|k}$ that appears on the diagonal of ${\bm P}^i_{k|k}$.  We can re-state the expressions of the previous section in  scalar form. Prior to receiving an update, the navigation uncertainty evolves according to \eqref{eq:Covariance Prediction_nomsmt}, which is  equivalent to 
\begin{equation*}
\nu^i_{k|k} = \nu^i_{k-1|k-1}  + \nu_w   \;.
\end{equation*}
If the $i$th agent receives a measurement at timestep $Z$, and the CNA has navigation uncertainty $\nu^c_{Z|Z}$ at this time, then the measurement noise variance  is 
\begin{equation*}
r_Z = \nu_y + \nu^c_{Z|Z}\;.
\end{equation*}
The covariance conditioned on the timestep prior is 
\begin{align*}
    \nu^i_{Z|Z-1} &=\nu^i_{Z-1|Z-1} + \nu_w =\nu^i_{0|0} +  Z  \nu_w \;,
\end{align*}
and the innovation covariance and Kalman gain are
\begin{align*}
    s_Z &= r_Z + \nu^i_{Z|Z-1}  = ( \nu_y + \nu^c_{Z|Z}) + (\nu^i_{0|0} + Z \nu_w) \\
    k_Z &= \nu^i_{Z|Z-1}/s_Z  \;.
\end{align*}
The posterior estimate variance at time $Z$ after the measurement is processed is:
\begin{align}
    \nu^i_{Z|Z}  &=  (1 - k_Z)\nu^i_{Z|Z-1} 
\\ 
    &=  \left(1 - \frac{[ \nu^i_{0|0} + Z\nu_w]}{( \nu_y + \nu^c_{Z|Z}) + (\nu^i_{0|0} + Z\nu_w)}\right)  [ \nu^i_{0|0} + Z \nu_w]  \;.
\label{eq:vZZ}
\end{align}
No additional measurements are obtained from times $Z+1$ to $T$ since we assume the CNA intercepts each agent only once and provides a single measurement. Then the uncertainty after the intercept grows at the same rate as before. 
In summary, the variance of the $i$th agent is 
\begin{align}
    \nu^i_{k|k} &= 
\begin{cases}
\nu^i_{0|0} + k \nu_w  &  0 \leq k \leq Z-1 \\
 \nu^i_{Z|Z}  + (k-Z)\nu_w   & Z \leq k \leq T 
\label{eq:vkk}
\end{cases}
\;.
\end{align}
\subsection{CNA Navigation Uncertainty after Surfacing}
Prior to starting a  surface task i.e., for times $k \leq S$, the CNA has variance
\begin{align*}
    \nu^c_{k|k} = \nu_G  + k \nu_c \;.
\end{align*}
$S+M \leq T$ is the time at which the vehicle is assumed to complete surfacing. Once it surfaces the variance  becomes $\nu^c_{S+M|S+M} = \nu_G$. 
Afterwards, for $k > S+M$, the CNA navigation uncertainty is
\begin{align*} 
    \nu^c_{k|k} = \nu_G + k \nu_c \;.
\end{align*}
In summary, the variance of the CNA is
\begin{align}
    \nu^c_{k|k} &= 
\begin{cases}
\nu_G + k \nu_c  &  0 \leq k < S+M \\
\nu_G + (k - (S+M)) \nu_c  & S+M \leq k \leq T
\end{cases}
\label{eq:nuc_kk} \;.
\end{align}
\subsection{Optimal Time-to-Aid}
\begin{figure}
\centering
\includegraphics[width=0.485\textwidth]{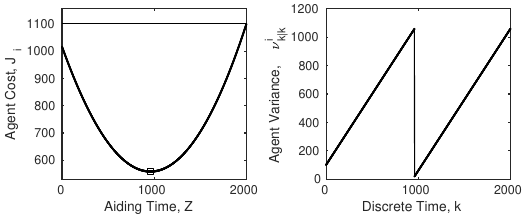} \\
\includegraphics[width=0.485\textwidth]{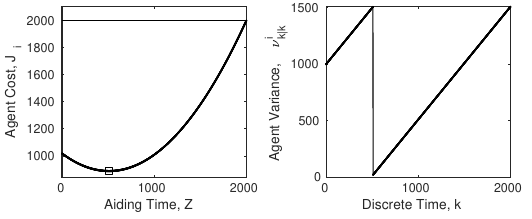} \\
\includegraphics[width=0.485\textwidth]{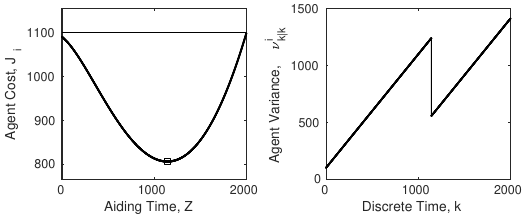} 
\caption{Agent cost function and agent variance with time for the case of $(\nu_{0|0}^i,\nu^c_{Z|Z}) = (100,10)$ (top), $(\nu_{0|0}^i,\nu^c_{Z|Z}) = (1000,10)$ (middle), $(\nu_{0|0}^i,\nu^c_{Z|Z}) = (100,1000)$ (bottom) assuming that $T = 2000$. The optimal time-to-aid, as computed by \eqref{eq:zstar}, is indicated by a square marker in each case. }
\label{fig:examples}
\end{figure}

As described above, the evolution of the agent's navigation uncertainty depends on when it receives a measurement $Z$. 
The cost \eqref{eq:prob} can be rewritten using the scalar Kalman filter notation as 
\begin{equation}
J' = \frac{1}{2}J = \frac{1}{N(T+1)} \sum^N_{i=1} \sum^{T}_{k=0}\nu^i_{k|k} = \frac{1}{N} \sum_{i=1}^N J_i \;,
\label{eq:Cost}
\end{equation}
where $J_i$ is the  average navigation uncertainty for the $i$th agent. $J_i$ can be computed by averaging \eqref{eq:vkk} where $ \nu^i_{Z|Z}$ is given in \eqref{eq:vZZ} and depends on  $\nu^c_{Z|Z}$ via \eqref{eq:nuc_kk}. For now,  suppose that $\nu^c_{Z|Z}$  is fixed. 
The cost $J_i$ is then computed by substituting \eqref{eq:vkk} into \eqref{eq:Cost}
\begin{align}
J_i(Z) 
=& \frac{1}{T+1}\sum_{k=0}^{T} \nu^i_{k|k}  \notag \\
=& \frac{1}{T+1}\left[\sum_{k=0}^{Z-1} (\nu^i_{0|0} + k \nu_w) + \sum_{k=Z}^{T} (  \nu^i_{Z|Z}  + (k-Z)\nu_w  )\right]  \notag \\
=& \frac{\nu_w T}{2} + \notag \\
& \frac{ Z \nu^i_{0|0}  +   (\nu^i_{Z|Z} - Z  \nu_w) (T+1) -  \nu^i_{Z|Z}Z  +   \nu_w Z^2  }{T+1}  \;.
\label{eq:J_i}
\end{align}
Note that $v_{Z|Z}^i$ is given in \eqref{eq:vZZ} and \eqref{eq:J_i} is a cubic in $Z$ (see Fig.~\ref{fig:examples}). 
To determine the optimal time-to-aid, $Z^*$, take the derivative of \eqref{eq:J_i} with respect to $Z$ to find the critical points:
\begin{align*}
\frac{\partial J_i}{ \partial Z} &=\frac{2 \nu_w Z  + (\nu_{0|0}^i - \nu_w T)}{T+1} + \frac{\partial(v_{Z|Z}^i/\partial Z)(T-Z)- v_{Z|Z}^i}{T+1}
\end{align*}
One of the solutions of ${\partial J_i}/{ \partial Z} = 0$ is a positive $Z^*$ corresponding to the optimal time-to-aid 
\begin{align}
Z_i^*= \frac{ \alpha +(T+1) \nu_w-3 \nu_{0|0}^i-3 \beta}{4 \nu_w}\label{eq:zstar} \;, 
\end{align}
\text{where}
\begin{align*}
\alpha &= \sqrt{(T \nu_w+\nu_{0|0}^i+\nu_w+\beta) (T \nu_w+\nu_{0|0}^i+\nu_w+9\beta)} \\
\beta &= \nu_y + \nu_{Z|Z}^c \;.
\end{align*}
\subsection{Bounds on the Cost Function}
In the absence of any aiding, the cost  for each agent reaches a maximum value
\begin{align}
J_i^{\rm max}
&= \frac{1}{T+1}\left[\sum_{k=0}^{T} (\nu^i_{0|0} + k \nu_w) \right] = \nu^i_{0|0}+ \nu_wT/2 %
\label{eq:Jimax} \;.
\end{align}
Recall, that the optimized $Z_i^*$ assumed the CNA had a fixed value $\nu_{Z|Z}^c$. Due to the choice of initial conditions and the CNA's surface task, $\nu_{Z|Z}^c \geq \nu_G$. Thus, a best-case cost for the $i$th agent is 
\begin{align*}
J_i^{\rm min} = J_i(Z_i^*) 
\end{align*}
where $J_i(Z_i^*)$ is evaluated using \eqref{eq:J_i} with $\nu_{Z|Z}^c= \nu_G$. Thus, the cost function is bounded by:
\begin{align}
\sum_{i=1}^N J_i^{\rm min}   \leq J  \leq \sum_{i=1}^N  J_i^{\rm max}
\label{eq:bounds}
\end{align}
It is also useful to establish the bounds of the cost function part-way along its path (i.e., at time $k$ after the CNA has completed several tasks). It is evident from the previous examples that the cost function monotonically decreases for $Z \in \{1, \ldots, Z_i^* - 1 \}$, reaches a minimum at $Z_i = Z_i^*$, and then monotonically increases for $Z \in \{ Z_i^*+1, \ldots, T\}$. Thus, if the CNA has not yet aided agent $i$, then the cost $J_i$ at time $k$ is bounded by $J_i \geq  J_i^*(k)$ where
\begin{align}
J_i^{\rm min}(k) = 
\begin{cases}
J_i^{\rm min}& \text{if}~ k \leq Z_i^*   \\
J_i(k)  & \text{otherwise}
\end{cases} \;.
\label{eq:Jik}
\end{align}

\section{Sequential Aiding Algorithm}
\label{sec:optimization_methods}
The results in Sec.~\ref{eq:single_contact} established  the optimal time-to-aid, $Z_i^*$,  for each agent. However, reaching all of the agents at their optimal aiding times is not feasible. Instead, we propose a greedy algorithm that uses a heuristic to incorporate knowledge of $Z_i^*$  to sequentially select which agent to aid. The algorithm proceeds as follows:
\begin{enumerate}
\item Compute the worst-case cost $J_i^{\rm max}$ for each agent using \eqref{eq:Jimax}. Let $C= \{1, 2, \ldots, N\}$ store the list of agents not yet visited. Initialize an empy task sequence ${A} \gets \emptyset$. 
\item While $|{A}| < {\rm min}(D,N)$ and $|C| > 0$ do the following:
\begin{enumerate}
\item For all agents $i \in C$:
\begin{enumerate}
\item Plan an intercept trajectory and record the time-to-intercept $\tau_i$ from \eqref{eq:intercept_time} and its discrete-time equivalent $Z_i^{\rm cand}$. 
\item Compute  the cost $J_i(Z_i^{\rm cand})$ if agent $i$ is selected for the next task using \eqref{eq:J_i}.
\item Compute the reward for selecting agent $i$:
\begin{equation}
r_i = \alpha \Delta^{\rm max}_i - \beta \Delta^{\rm opt}_i  - \gamma (\tau_i/T) \;,
\label{eq:reward}
\end{equation}
where
\begin{align*}
\Delta^{\rm max}_i &= [J_i^{\rm max} - J_i(Z_{\rm cand})]/J_i^{\rm max}  \\
\Delta^{\rm opt}_i &= [J_i^*(k)- J_i(Z_{\rm cand})]/J_i(Z_i^{\rm cand})\;.
\end{align*}
\end{enumerate}
\item Select the agent $ i^*\in C$ that maximizes the reward  $r_i $. Update the elapsed mission time $T$. If $ T \leq T_{\rm max}$, then add agent $i^*$ to ${A}$ and remove it from $C$. Otherwise,  remove $ i^*$ from $C$.
\item Update the CNA position and elapsed time.
\end{enumerate}
\item If $D = N+1$ then the current task sequence ${A}$ has $N$ elements and is missing one task. Compare inserting  the surfacing task at each position in ${A}$  and return the lowest cost sequence ${A}^*$ according to \eqref{eq:Cost}. 
\item If $D \leq N$ then the current task sequence ${A}$ cannot be enlarged. Compare replacing each agent task with a  surfacing task and return the task sequence ${A}^*$ that yields the lowest cost  \eqref{eq:Cost}. 
\end{enumerate}
The above greedy algorithm selects the highest reward at each step in the task allocation. The reward \eqref{eq:reward} is designed to balance three objectives: (1)  to secure a large reduction in the overall cost by selecting a large $\Delta^{\rm max}_i $, (2) to avoid a penalty for aiding an agent at a sub-optimal time (i.e., far away from its optimal time-to-aid as quantified by $\Delta^{\rm opt}_i $), and (3) to avoid long transit times $\tau_i$. Each of the three terms in \eqref{eq:reward} have similar magnitude and are scaled by user-selected weights  $(\alpha, \beta, \gamma)$.

The greedy heuristic algorithm described above is efficient to compute, but is suboptimal. For a sufficiently large $N$, all of the candidate task sequences in $\mathcal{F}$ can be enumerated and exhaustively compared to find the optimal task sequence --- we refer to this approach as the optimal (exhaustive enumeration) algorithm.

\section{Numerical Simulations}
\label{sec:results}
This section presents an illustrative example and statistical results obtained by a randomized Monte Carlo experiment comparing the greedy and optimal algorithms. 
\subsection{Illustrative Example}
\begin{figure}
\centering
\includegraphics[width=0.37\textwidth]{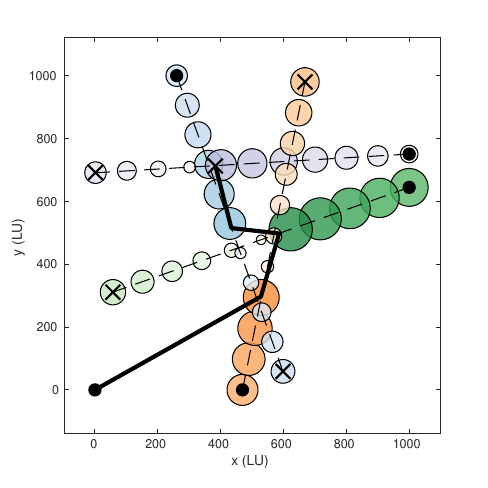}
\includegraphics[width=0.37\textwidth]{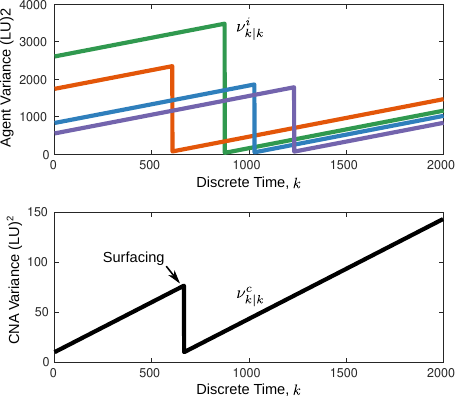}
\caption{Top: Trajectories of agents (dashed line) along with their starting locations (circular markers) and end locations (``x" markers) with an overlay of an uncertainty circle indicating $\nu_{k|k}^i$ along the path. The optimized path of the CNA is shown as the solid black line. Middle and bottom: navigation variance as a function of time for the agents and CNA. Discontinuities in $\nu_{k|k}^i$ indicate an aiding measurement was received by the agent. The discontinuity in $\nu_{k|k}^c$ indicates the CNA completed a surfacing.}
\label{fig:example}
\end{figure}

A scenario was simulated that consisted of $N=4$ agents with initial positions and heading as shown in Fig.~\ref{fig:example}. The CNA was initialized at the origin. The simulation used the parameters in Table~\ref{tb:parameters} and the agents were initialized with random selected initial variance of $\nu_{0|0}^i = \{2610, 1747, 837, 557 \}$ for agents $i =1, 2, 3$ and $4$, respectively.
\begin{table}
\begin{center}
\caption{Parameters used in numerical simulations. LU = length unit, TU = time unit.}\label{tb:parameters}
    \begin{tabular}{|c|c|c|}
    \hline
        {\bf Parameter} & {\bf Symbol} & {\bf Value} \\ 
\hline
Number of steps for surface & $M$ & 60 \\
Mission time & $T_{\rm max}$ & 2000 TU \\
Timestep & $\Delta t$ & 1 TU \\
Speed of CNA & $v_{\rm max}$ & 1 LU/TU\\
Speed of Agents & $v_a$ & 0.5 LU/TU \\
Process variance of agents & $\nu_w$ & 1 LU$^2$\\
Process variance of CNA & $\nu_c$ & $\nu_w/10$ LU$^2$\\
Initial variance of agents & $\nu^i_{0|0}$ & LU$^2$\\
Measurement noise & $\nu_y$ & 10 LU$^2$\\
Variance of CNA upon GPS update & $\nu_G$ & 10 LU$^2$\\
\hline
\end{tabular}
\end{center}
\end{table}
Numerical simulations display the 50\% confidence ellipse for the uncertain position of the agents. 
The CNA path was solved for using the optimal (exhaustive enumeration) algorithm to give the task sequence ${A} =\{2,   0,     1,     3,     4\}$. That is, the CNA first aids agent 2 (orange), then surfaces, then continues to aid agents 1 (green), 3 (blue), and 4 (purple). The overall cost of this task sequence  was $J=1172$.  Even though the agent  1 had the largest initial variance, the algorithm chooses to aid agent 2 first. Intuitively, agent 2 is closer to the CNA starting location and it more efficient to aid them along the way to agent 1. The CNA reached the final agent 4 at time $k = 1233$. A modified problem formulation that allows the CNA to revisit agents with remaining mission time may be considered in future work.  

\subsection{Monte Carlo Simulation}
The greedy algorithm was  tested through a Monte Carlo simulation that randomized the initial locations, heading angles, and initial variance  of the agents according to the following three strategies:
\begin{enumerate}
\item Agents were randomly position in the interior of a box with side-length $L = 1000$ and a random heading.
\item Agents were randomly positioned on the boundary of the box described above with a heading that points 
towards the center of the box and is perturbed by an angle randomly drawn from $[-30,30]$ deg.  
\item Agents were randomly positioned in a circular formation with a center randomly chosen on the boundary of the box and synchronized heading angles that point towards the center of the box. 
\end{enumerate}
The number of agents was varied from $N =3$ to $N=14$ and for each $N$ value 100 random scenarios were generated. The parameters in the simulation are listed in Table~\ref{tb:parameters}. 
Approximately 1/3 of each set of trials adopted each of the three initialization strategies described above and the initial variance of the agents was randomly chosen from the interval $\nu_{0|0}^i \in (0, 3000]$. The greedy heuristic search algorithms were evaluated with the following combination of weights:  G1 $= (\alpha, \beta, \gamma) = (1,0,0)$
, G2 $= (\alpha, \beta, \gamma) = (0,1,0)$
, G3 $= (\alpha, \beta, \gamma) = (0,0,1)$
, and G4 $= (\alpha, \beta, \gamma) = (1,0.5,0.5)$.
This  allowed comparing the influence of each term in the reward \eqref{eq:reward}, as well as their combined effect.
With $N= 10$ agents each trial took about 10 minutes to solve using the optimal algorithm, thus evaluations with this algorithm were limited to $N  \leq 10$. Since the optimal algorithm tests all possible task sequences, the worst-case choice of task sequence was also  recorded for comparison. The algorithms were all implemented in MATLAB using a parallelized computation on a laptop computer. 

\begin{figure}
    \centering
    \includegraphics[width = 0.48\textwidth]{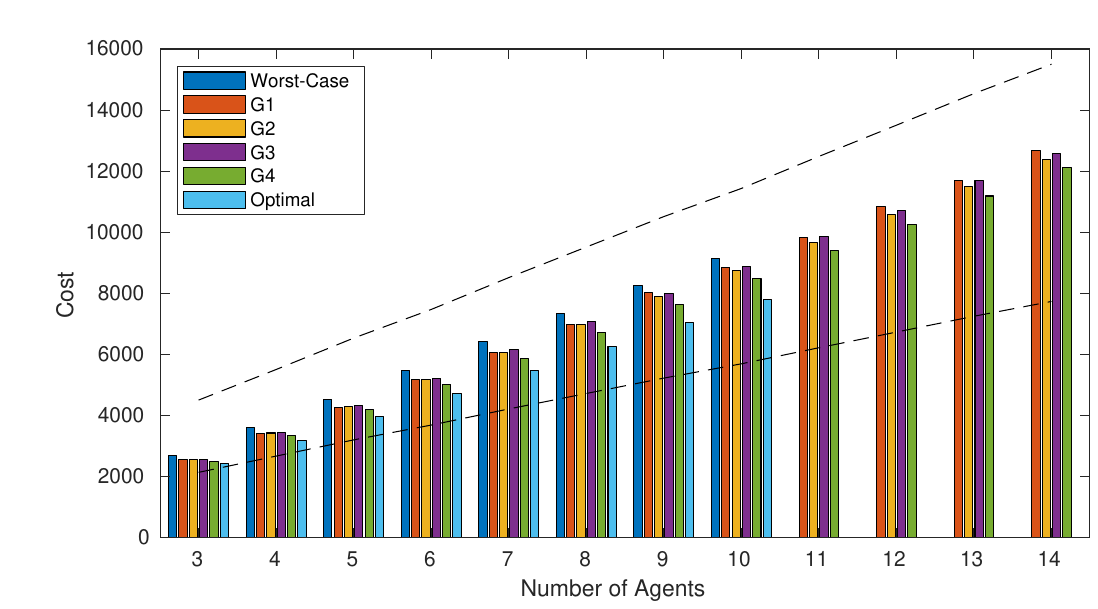} \\
    \includegraphics[width = 0.48\textwidth]{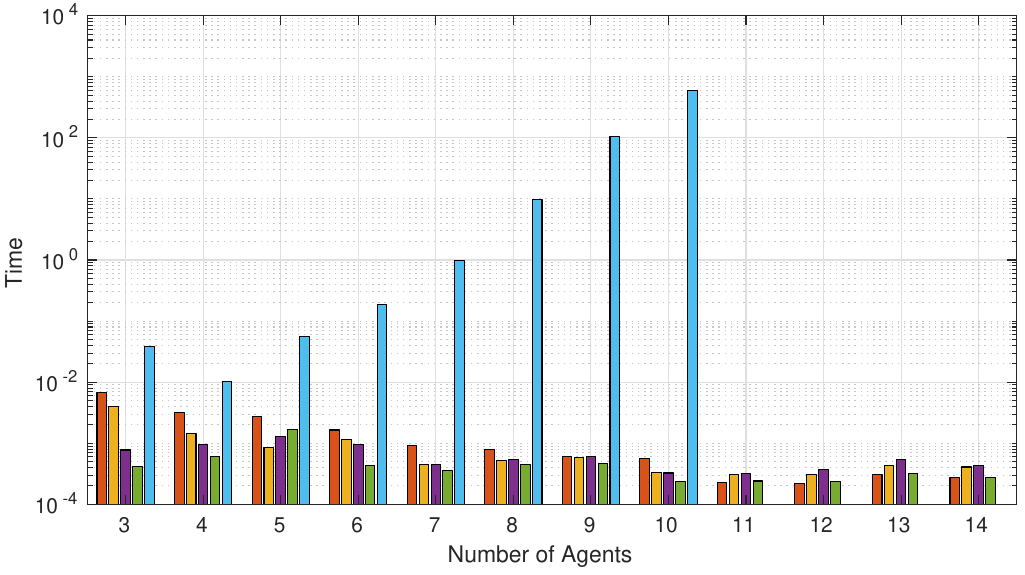}
    \caption{Monte Carlo simulation results. The cost bounds from \eqref{eq:bounds} are shown as dashed lines in the top panel.}
    \label{fig:monte_carlo}
\end{figure}

The mean cost and the mean computation time for each algorithm is shown in Fig.~\ref{fig:monte_carlo}. By comparing the optimal and worst-case performance to the bounds of \eqref{eq:bounds}, it is apparent that the bounds bracket the performance but are not tight. When comparing the heuristic algorithms, the G4 set of $(\alpha, \beta, \gamma)$ parameters that mix all three terms in \eqref{eq:reward} outperforms G1, G2, G3, which each only use one of the terms in \eqref{eq:reward}. While the computation time of the optimal (exhaustive) algorithm increases exponentially, the average computation time of the greedy algorithms remains below $0.01$ seconds. Unexpectedly, the computation time increases slightly for a smaller number of agents --- this may be due to how the experiment was implemented with the parallel processing toolbox in MATLAB. 

\section{Conclusion}
\label{sec:conclusion}
This paper considered the problem of planning a planar path for a CNA to sequentially provide navigation aiding measurements to a set of $N$ agents while also considering the possibility of surfacing to improve its own navigation uncertainty. 
A greedy algorithm was formulated to solve for the desired CNA path by considering a heuristic that  considers three weighted factors: (1) selecting agents  that lead to large reductions in cost, (2) selecting agents for whom the aiding time is close to the optimal time-to-aid, and (3) selecting agents that are close to the CNA. A Monte Carlo experiment comparing the greedy algorithm to an optimal algorithm showed that by combining all three weighted factors a lower cost path is returned than by considering each factor independently. 

Future work may consider heuristics with recipes for choosing various  tuning parameters or other optimization schemes. Other scenarios could also be investigated---for example, a CNA that can revisit agents and surface multiple times, a CNA that provides water current estimates to agents to minimize final position errors, or a CNA with a finite communication radius that can service multiple agents simulatenously. Aiding agents with unknown trajectories (i.e., requiring an initial search to localize them) may also be of interest.
\bibliography{refs} 
\end{document}